%% file: hierarchical_code_summarization.tex
\documentclass{article}

\usepackage{PRIMEarxiv}

\usepackage[utf8]{inputenc} 
\usepackage[T1]{fontenc}    
\usepackage{hyperref}       
\usepackage{url}            
\usepackage{booktabs}       
\usepackage{amsfonts}       
\usepackage{nicefrac}       
\usepackage{microtype}      
\usepackage{graphicx}
\usepackage{amsmath}
\usepackage{amssymb}
\usepackage{mathtools}
\usepackage{amsthm}
\usepackage{listings}
\usepackage{xcolor}
\usepackage{algorithm}
\usepackage{algpseudocode}
\usepackage{datatool}
\usepackage{caption}
\usepackage{float} 

\definecolor{codegreen}{rgb}{0,0.6,0}
\definecolor{codegray}{rgb}{0.5,0.5,0.5}
\definecolor{codepurple}{rgb}{0.58,0,0.82}
\definecolor{backcolour}{rgb}{0.95,0.95,0.92}

\lstdefinestyle{mystyle}{
  backgroundcolor=\color{backcolour},
  commentstyle=\color{codegreen},
  keywordstyle=\color{magenta},
  numberstyle=\tiny\color{codegray},
  stringstyle=\color{codepurple},
  basicstyle=\ttfamily\footnotesize,
  breakatwhitespace=false,
  breaklines=true,
  captionpos=b,
  keepspaces=true,
  numbers=left,
  numbersep=5pt,
  showspaces=false,
  showstringspaces=false,
  showtabs=false,
  tabsize=2
}

\date{}

\title{\Large \bf Code-Craft: Hierarchical Graph-Based Code Summarization for Enhanced Context Retrieval}

\author{
{\rm David Sounthiraraj, Jared Hancock, Yassin Kortam } \vspace{1mm} \\
{\rm Ashok Javvaji, Prabhat Singh, Shaila Shankar } \vspace{1mm} \\
{\rm Cisco Systems} \vspace{1mm} \\
{\rm \texttt{dsounthi, jarhanco, ykortam}} \vspace{1mm} \\
{\rm \texttt{javvaji, prabhat7, shailas}}
}

\begin{document}
\maketitle

\thispagestyle{empty} 

\begin{abstract}
Understanding and navigating large-scale codebases remains a significant challenge in software engineering. 
Existing methods often treat code as flat text or focus primarily on local structural relationships, limiting their ability to 
provide holistic, context-aware information retrieval. We present \emph{Hierarchical Code Graph Summarization (HCGS)}, a novel
approach that constructs a multi-layered representation of a codebase by generating structured summaries in a bottom-up fashion 
from a code graph. HCGS leverages the Language Server Protocol for language-agnostic code analysis and employs a parallel level-based 
algorithm for efficient summary generation. Through extensive evaluation on five diverse codebases totaling 7,531 functions, 
HCGS demonstrates significant improvements in code retrieval accuracy, achieving up to 82\% relative improvement in top-1 retrieval 
precision for large codebases like libsignal (27.15 percentage points), and perfect Pass@3 scores for smaller repositories. 
The system's hierarchical approach consistently outperforms traditional code-only retrieval across all metrics, with particularly 
substantial gains in larger, more complex codebases where understanding function relationships is crucial.
\end{abstract}

\section{Introduction}

Understanding and navigating large-scale codebases remains a significant challenge in software engineering. A comprehensive systematic review by Liu et al. \cite{Liu_2021} reveals that despite decades of research, effective code search and comprehension tools remain elusive. While Large Language Models (LLMs) have shown proficiency in understanding localized code snippets, they struggle to grasp the full scope of entire codebases due to their limited context windows, which hinders their ability to navigate the complex interdependencies inherent in large software systems. Ma et al. \cite{ma2024understand} emphasizes that comprehensive repository-level understanding is critical for effective automated software engineering, identifying key challenges including the impracticality of including thousands of files in LLM context windows and the difficulty of identifying code relevant to specific issues within extensive contexts.

Code graphs have emerged as a promising approach to represent structural relationships between code elements \cite{Allamanis2018Learning, codegraph}, providing valuable architectural insights and enhancing visualization of code relationships. Liu et al. \cite{Liu_2021} notes that while 51\% of contemporary code search tools focus on text-based approaches, only a small fraction effectively leverage structural information. Recent work by Liu et al. \cite{liu2024graphcoderenhancingrepositorylevelcode} demonstrated the effectiveness of structured code representations through their GraphCoder framework, which uses a Code Context Graph combining control flow and dependency relationships to enhance repository-level code completion. Despite these advances, existing graph-based approaches often lack the capability to abstract details at varying levels of granularity, impeding high-level codebase understanding.

The integration of structural insights from code graphs with the summarization capabilities of language models presents a promising avenue for addressing the limitations inherent in each approach when used independently. This aligns with the emerging trend identified by Liu et al. \cite{Liu_2021}, where deep learning has become the most popular modeling technique for code understanding. Jain et al. \cite{jain2024llmagentsimprovesemantic} demonstrated how agentic LLMs combined with Retrieval Augmented Generation (RAG) can significantly improve semantic code search by addressing the "vocabulary mismatch problem," achieving a 78.2\% success rate at retrieving relevant code snippets within the top 10 results. Similarly, Ma et al. \cite{ma2024understand} introduced RepoUnderstander, employing a top-down approach to condense repository information into a knowledge graph and using Monte Carlo Tree Search to efficiently explore repositories, achieving a 21.33

Systematic analysis reveals several persistent challenges in code understanding and retrieval \cite{Liu_2021}, including limited codebase diversity, query limitations, model construction issues, and evaluation challenges, with 55\% of tools still relying on subjective manual evaluation methods. The vocabulary mismatch problem \cite{jain2024llmagentsimprovesemantic} and information asymmetry in benchmark datasets \cite{ma2024understand} further complicate effective code search and understanding. These challenges highlight the need for more robust, multi-language approaches with improved evaluation methodologies and context-aware search capabilities.

We introduce Hierarchical Code Graph Summarization (HCGS), a system that addresses these limitations by constructing a structured, hierarchical representation of a codebase. HCGS employs a bottom-up traversal strategy on a code graph, generating structured summaries at each level. This approach ensures that higher-level summaries are informed by the summaries of their constituent components, creating a rich, multi-layered understanding of the code. Our key contributions are: (1) A novel hierarchical summarization approach based on bottom-up traversal of code graphs; (2) A structured summary schema that captures both implementation details and dependencies; and (3) Vector-based retrieval on the structured summaries.

\section{Related Work}
\label{sec:related_work}

Research on code understanding and retrieval spans multiple interconnected domains. Traditional code search tools rely on keyword matching and regular expressions, while more advanced techniques leverage abstract syntax trees \cite{black2002abstract} or code graphs \cite{waddell2007practical}. Liu et al.'s systematic review \cite{Liu_2021} reveals that 74\% of code search tools employ natural language queries, highlighting the importance of bridging the semantic gap between natural language and code structure, though 55\% still rely on subjective manual evaluation methods. Deep learning has emerged as the dominant approach in recent years \cite{Gu2018DeepCS, Haldar2020}, particularly for addressing the semantic gap between queries and code.

Recent advancements have leveraged agentic LLMs and Retrieval Augmented Generation to enhance search accuracy. Jain et al. \cite{jain2024llmagentsimprovesemantic} introduced RepoRift, which uses specialized LLM agents to recursively augment user queries with relevant context, addressing the "vocabulary mismatch problem" by injecting technical details into queries. Their multi-stream ensemble approach achieved a 78.2\% success rate at Success@10 on the CodeSearchNet dataset, significantly outperforming previous methods. Traditional code summarization techniques \cite{Allamanis2016Convolutional, Iyer2016Summarizing, Ahmad2020Transformer, LeClair2019Improved} typically focus on individual functions rather than entire codebases, often treating code elements in isolation and missing broader contextual relationships.

Static analysis techniques like control flow and data flow analysis \cite{allen1970control} provide valuable structural information but often produce low-level details that are difficult to interpret in large codebases and lack mechanisms for higher-level abstraction. Liu et al. \cite{Liu_2021} notes that heuristic models based on program analysis techniques require substantial domain knowledge to design effective matching algorithms.

More recent approaches represent codebases as multi-level graphs. CodexGraph \cite{codexgraph2024} and RepoHyper \cite{yang2023repohyper} use code graphs to represent entire repositories with multiple levels of abstraction. Ma et al. \cite{ma2024understand} introduced RepoUnderstander, which constructs a repository knowledge graph in a top-down manner and employs Monte Carlo Tree Search to efficiently explore it, prioritizing critical information discovery. Liu et al. \cite{liu2024graphcoderenhancingrepositorylevelcode} developed GraphCoder, leveraging a statement-level Code Context Graph that combines control flow and dependency relationships, employing a coarse-to-fine retrieval process that significantly improved code completion accuracy while reducing retrieval time and storage requirements.

HCGS builds upon these prior works by combining code graphs, structured summarization, and efficient search techniques to create a hierarchical representation that facilitates both high-level understanding and detailed analysis. Unlike previous approaches, HCGS employs a bottom-up traversal strategy ensuring higher-level summaries inherit context from their constituent components, contrasting with RepoUnderstander's top-down approach. While RepoRift focuses on enhancing query context through LLM agents, RepoUnderstander emphasizes efficient repository exploration, and GraphCoder focuses on statement-level structural representation for code completion, HCGS complements these approaches by enhancing the code representation itself through hierarchical summarization. These complementary strategies address different aspects of repository-level understanding: HCGS focuses on rich contextual representation while RepoUnderstander emphasizes efficient exploration and information discovery. This approach directly addresses key challenges identified in the systematic review \cite{Liu_2021}, including the need for better code representations, improved learning models, and more effective evaluation methodologies.

\section{System Architecture}

\begin{figure}[htbp]
  \centering
  \includegraphics[height=0.38\textheight]{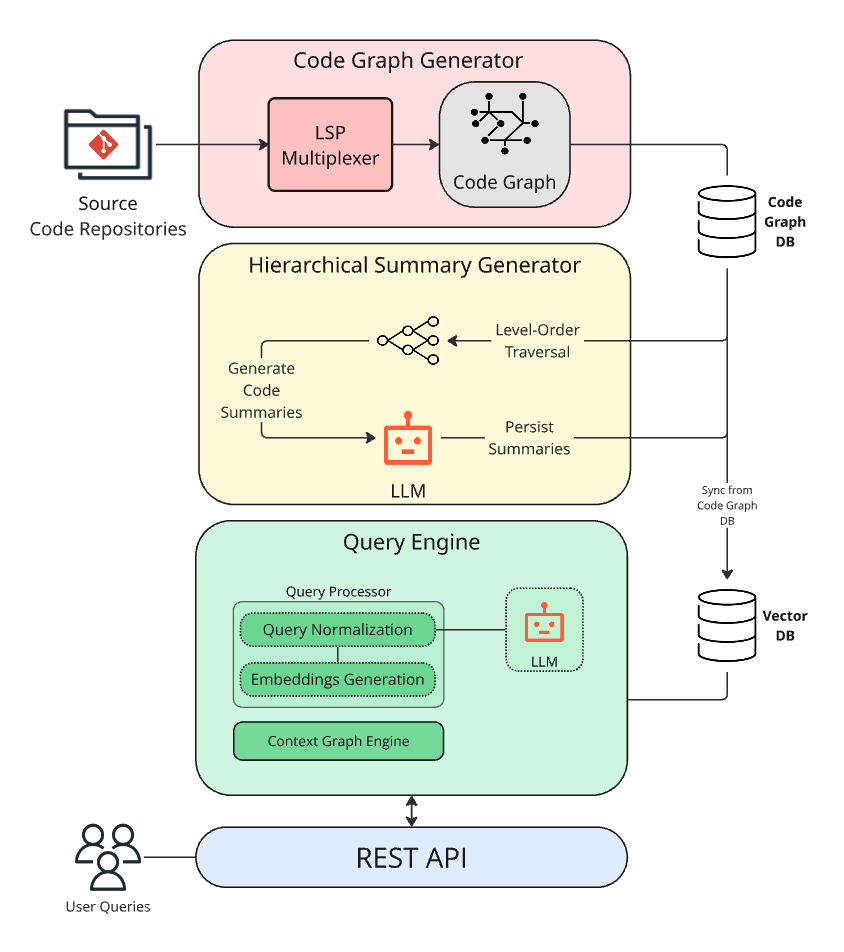} 
  \caption{HCGS System Architecture}
  \label{fig:architecture}
\end{figure}

HCGS consists of several primary components: a Code Graph Generator, a Summary Generator, a Storage layer, and a Query Engine. Figure~\ref{fig:architecture} illustrates the system architecture.

\vspace{0.5em} 

\vspace{0.5em} 

The Hierarchical Code Graph Summarization (HCGS) system addresses a fundamental limitation in traditional code embedding approaches. Conventional methods that generate embeddings for individual functions in isolation fail to capture the broader context in which these functions operate. This isolation leads to embeddings that lack critical information about the function's role within the larger system architecture.

\vspace{0.3em} 

\begin{figure}[htbp]
  \centering
  \includegraphics[width=0.95\textwidth]{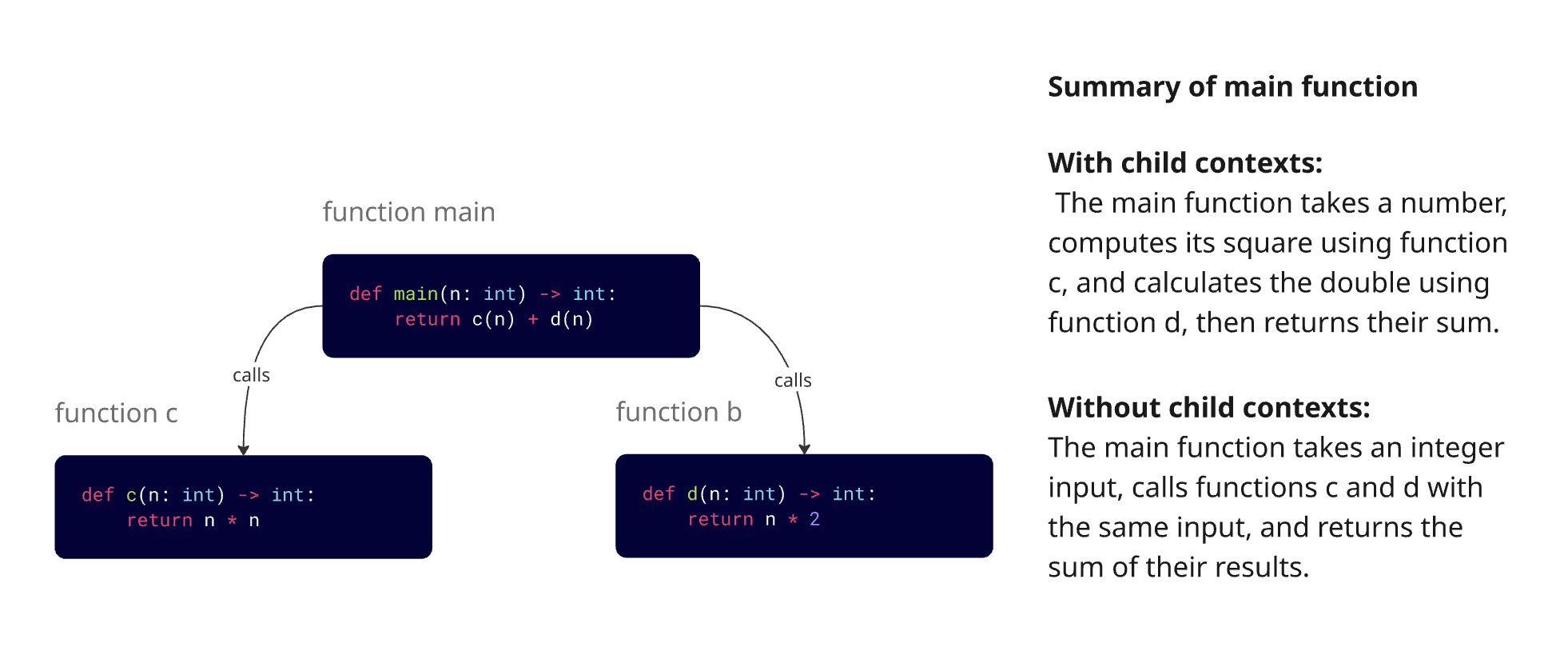} 
  \caption{Function Calls with Context}
  \label{fig:function_calls}
\end{figure}

\vspace{0.5em} 

As illustrated in Figure~\ref{fig:function_calls}, when considering a function in isolation, we miss the rich contextual information provided by its relationships with other functions. The hierarchical summarization approach captures this context by propagating information from called functions upward through the call graph. This results in embeddings with significantly higher information density and semantic accuracy compared to isolated function embeddings.

\vspace{0.8em} 

The HCGS approach leverages this contextual information to create a more comprehensive understanding of the codebase. By analyzing the relationships between functions and propagating information upward through the call graph, we can generate more accurate and informative embeddings that capture the true semantic meaning of the code.

\begin{figure}[htbp]
  \centering
  \includegraphics[width=0.92\textwidth]{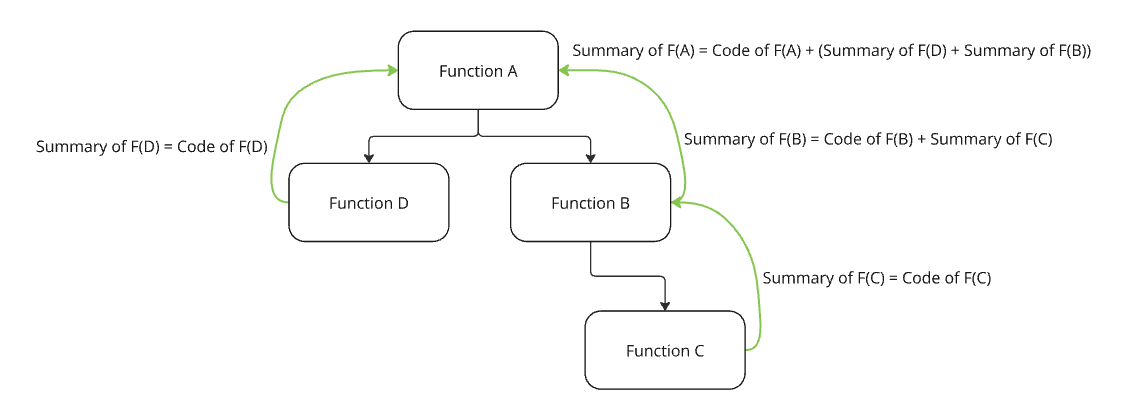} 
  \caption{HCGS Approach}
  \label{fig:hcgs_approach}
\end{figure}

\vspace{0.8em} 

Figure~\ref{fig:hcgs_approach} illustrates how our approach leverages advanced Large Language Models (LLMs) to generate hierarchical summaries that capture not only the syntactic structure of the code but also its semantic meaning and relationships. This approach is superior to traditional embedding methods that rely primarily on code structure, naming conventions, and comments, as it can understand and represent the deeper functional purpose and relationships within the codebase. The multi-level nature of these summaries enables a more comprehensive understanding of the code at different levels of abstraction.

\subsection{Code Graph Generator}

The Code Graph Generator takes the source code of a project as input and generates a directed graph where nodes represent code elements (e.g., files, classes, functions, methods) and edges represent relationships between them (e.g., calls, inheritance, imports). The graph is constructed using static analysis techniques. To achieve broad language support and robust parsing, we leverage the \textbf{Language Server Protocol (LSP)}. The LSP provides a standardized way to interact with language servers, which offer deep understanding of code structure and semantics for various programming languages.

For our LSP integration, we use a modified version of multilspy \cite{multilspy2023, agrawal2023guidinglanguagemodelscode}, a cross-platform library designed to simplify the process of creating language server clients. Multilspy was originally developed as part of research for Monitor-Guided Decoding of code language models \cite{agrawal2023guidinglanguagemodelscode}. The library enhances our system by handling automated server management for language-specific binaries, abstracting the JSON-RPC communication protocol, providing a unified interface for multiple language servers, and enabling essential static analysis operations such as finding definitions, references, and symbol information.

By utilizing the LSP through multilspy, the HCGS system gains several key capabilities. It provides \textbf{Abstract Syntax Trees (ASTs)} for a given file from the language server, which forms the foundation for our code graph, providing a structured representation of the code that preserves syntactic relationships.

We're able to accurately identify and resolve dependencies between code elements (e.g., function calls, class inheritance) by requesting information about symbols, definitions, and references from the LSP. This dependency information is crucial for constructing the hierarchical relationships in the code graph.

The ability to support multiple languages is simple, as it is only a matter of invoking the appropriate language server for each language. This eliminates the need for language-specific parsers and allows HCGS to be deployed across heterogeneous codebases that combine multiple programming languages.

Language servers are typically maintained by language experts and are constantly updated to handle language changes and edge cases, leading to more accurate and robust code graph construction compared to custom parsing solutions that may not stay current with language evolution. 
In this regard, the LSP acts as a powerful abstraction layer that insulates HCGS from the complexities of individual programming languages, allowing it to focus on higher-level code understanding tasks.

For graph storage, HCGS employs a structured summary schema to represent the information extracted from each code element. This schema facilitates consistent and comprehensive documentation while enabling efficient, targeted querying.

\begin{figure}[htbp]
  \centering
  \includegraphics[width=0.95\textwidth]{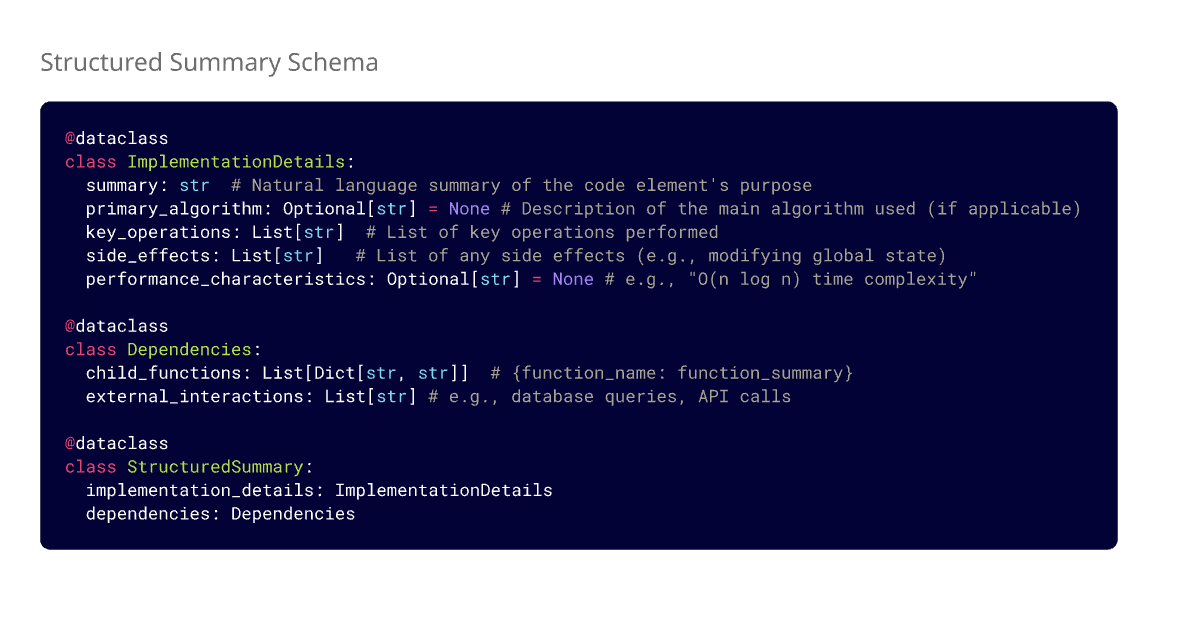} 
  \caption{Structured Summary Schema}
  \label{fig:schema}
\end{figure}

\vspace{0.5em} 

As shown in Figure~\ref{fig:schema}, this structured schema is crucial for enabling the hierarchical summarization process and efficient retrieval, as it organizes code information into well-defined categories including implementation details and dependencies.

\subsection{Summary Generator}
\label{sec:summary_generator}

The Summary Generator is responsible for creating structured summaries of code elements based on the information extracted from the code graph. For each node in the graph, HCGS issues a prompt
to an LLM to generate the summaries. For this paper, we used the Claude Haiku 3.5 model\footnote{https://www.anthropic.com/claude/haiku}.

\begin{figure}[!ht]
  \centering
  \includegraphics[width=0.8\textwidth,height=0.5\textwidth,keepaspectratio]{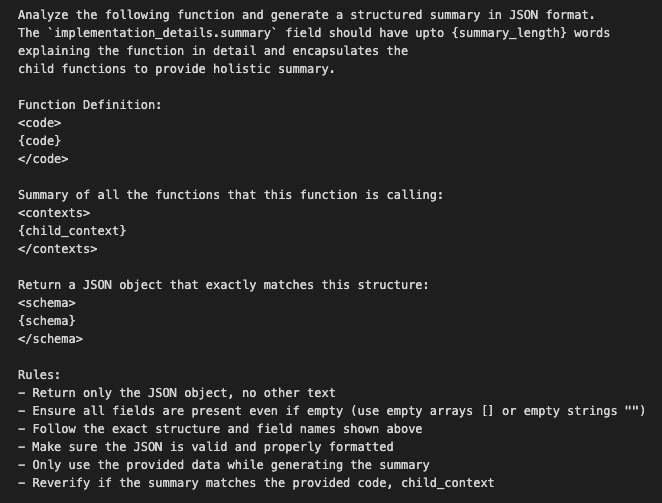}
  \caption{Summarization Prompt Template}
  \label{fig:prompt_template}
\end{figure}

Figure~\ref{fig:prompt_template} shows the actual prompt template used by our system to generate code summaries. This carefully engineered prompt instructs the LLM to analyze both the function's source code and the summaries of any called functions, then generate a structured summary following our schema. 
The hierarchical nature of our approach is evident in how the template incorporates child function summaries via the \texttt{child\_context} parameter, ensuring that higher-level summaries are contextually aware of their dependencies.

Two distinct algorithms for generating hierarchical summaries: a sequential DFS-based approach and a parallel level-based approach. 
The level-based algorithm represents a significant advancement over the traditional DFS-based approach, offering improved scalability and performance for large codebases.

The key innovation of the level-based approach is the organization of the code graph into dependency levels, where each level contains functions that can be processed 
independently once all functions in lower levels have been processed. This enables efficient parallel processing while maintaining deterministic behavior.

\subsubsection{Level Order Traversal Construction}

The first step is organizing the function dependency graph into levels, as shown in Algorithm~\ref{alg:level_construction}:

\vspace{-0.5em} 
\begin{algorithm}[H]
\caption{Level Construction}\label{alg:level_construction}
\begin{algorithmic}[1]
\Function{BuildLevels}{$nodes, graph$}
  \State $deps \gets \{\}$ \Comment{Dependency map}
  \State $levels \gets []$ \Comment{List of levels}

  \For{$id \text{ in } nodes$}
    \State $deps[id] \gets graph.get\_edges(id)$
  \EndFor

  \State $remaining \gets Set(nodes)$
  \While{$remaining \neq \emptyset$}
    \State $level \gets []$
    \State $to\_remove \gets []$

    \For{$id \text{ in } remaining$}
      \If{$AllDepsProcessed(id, deps, remaining)$}
        \State $level.append(id)$
        \State $to\_remove.append(id)$
      \EndIf
    \EndFor

    \If{$to\_remove.isEmpty()$}
      \State $id \gets BreakCycle(remaining, deps)$
      \State $level.append(id)$
      \State $to\_remove.append(id)$
    \EndIf

    \For{$id \text{ in } to\_remove$}
      \State $remaining.remove(id)$
    \EndFor

    \State $levels.append(level)$
  \EndWhile

  \State \Return $levels$
\EndFunction
\end{algorithmic}
\end{algorithm}

This algorithm transforms the function dependency graph into levels where each level contains functions that only depend on functions in lower levels. It handles cyclic dependencies by strategically breaking cycles when necessary.

\subsubsection{Parallel Level Processing}

Once the levels are constructed, each level is processed in parallel as shown in Algorithm~\ref{alg:parallel_processing}:

\vspace{-0.5em} 
\begin{algorithm}[H]
\caption{Parallel Processing}\label{alg:parallel_processing}
\begin{algorithmic}[1]
\Procedure{ProcessLevels}{$levels, graph$}
  \State $cache \gets \{\}$ \Comment{Summary cache}
  \State $modules \gets \{\}$ \Comment{Affected modules}

  \For{$level \text{ in } levels$}
    \State $results \gets \{\}$
    \State $futures \gets \{\}$
    \For{$id \text{ in } level$}
      \State $futures[id] \gets ThreadPool.submit(SummarizeNode, id, cache)$
    \EndFor

    \For{$id, future \text{ in } futures.items()$}
      \State $summary \gets future.result()$
      \If{$summary \neq null$}
        \State $results[id] \gets summary$
        \State $cache[id] \gets summary$
        \State $node \gets graph.get\_node(id)$
        \State $file \gets node.file\_path$
        \If{$file \notin modules$} \Comment{Track Module}
          \State $modules[file] \gets []$
        \EndIf
        \State $modules[file].append(node)$
      \EndIf
    \EndFor

    \For{$id, summary \text{ in } results.items()$}
      \State $Store(id, summary)$
    \EndFor
  \EndFor

  \State \Return $modules$
\EndProcedure
\end{algorithmic}
\end{algorithm}

This algorithm processes each level sequentially, but within each level, it processes all nodes in parallel using a thread pool. The cache ensures that child summaries are available when processing parent nodes.

\subsubsection{Node Summary Generation}

The core of the summary generation process:

\vspace{-0.5em} 
\begin{algorithm}[H]
\caption{Node Summary}\label{alg:node_summary}
\begin{algorithmic}[1]
\Function{SummarizeNode}{$id, cache, graph$}
  \State $node \gets graph.get\_node(id)$
  \State $child\_summaries \gets []$

  \For{$child\_id \text{ in } graph.get\_edges(id)$}
    \If{$child\_id \text{ in } cache$}
      \State $child\_summaries.append(cache[child\_id])$
    \EndIf
  \EndFor

  \State $summary \gets LLM.generate(node.code, child\_summaries)$
  \State \Return $summary$
\EndFunction
\end{algorithmic}
\end{algorithm}

This function collects summaries of child nodes and uses an LLM to generate a structured summary that incorporates both the node's code and the context from its dependencies.

\subsubsection{Module Summary Generation}

The final step aggregates function summaries into module summaries:

\vspace{-0.5em} 
\begin{algorithm}[H]
\caption{Module Summary}\label{alg:module_summary}
\begin{algorithmic}[1]
\Procedure{SummarizeModules}{$modules, graph$}
  \For{$file, nodes \text{ in } modules.items()$}
    \State $func\_summaries \gets []$
    \For{$node \text{ in } nodes$}
      \State $func\_summaries.append(\{$
      \State \quad $'name': node.name,$
      \State \quad $'summary': node.summary.details$
      \State $\})$
    \EndFor

    \State $module\_summary \gets LLM.generate(func\_summaries)$
    \State $module\_id \gets graph.get\_module\_id(file)$
    \State $Store(module\_id, module\_summary)$
  \EndFor
\EndProcedure
\end{algorithmic}
\end{algorithm}

This step completes the hierarchical representation by creating module-level summaries from function summaries.

The level-based approach offers several advantages over the traditional DFS-based method:
\begin{itemize}
  \item \textbf{Improved Parallelism:} By organizing functions into levels, the algorithm can process independent functions concurrently, significantly improving performance for large codebases.
  \item \textbf{Deterministic Behavior:} The level-based approach ensures that all dependencies are processed before a function, eliminating the need for complex cycle detection and backtracking.
  \item \textbf{Efficient Resource Utilization:} The algorithm processes one level at a time, leading to more predictable memory usage patterns and efficient resource utilization.
  \item \textbf{Enhanced Scalability:} The combination of parallelism and deterministic processing makes the algorithm highly scalable, capable of handling codebases with millions of lines of code.
\end{itemize}

This multi-stage approach creates a comprehensive hierarchical representation of the codebase, where each summary incorporates the context of its dependencies. The resulting summaries capture not only the syntactic structure of the code but also its semantic meaning and relationships, providing a rich foundation for subsequent retrieval and analysis tasks.

\vspace{1em} 

\subsection{Storage and Retrieval}
\vspace{0.3em} 



While other RAG methods may prioritize explicit code structure\cite{codexgraph2024}, the hierarchical summaries provide a rich semantic meaning in addition to a graph-like expression of relationships to 
other code elements inside and outside of its repository. HCGS generates \textbf{Dense Vector Embeddings} from the structured summaries and persists them in a vector database\footnote{ChromaDB \url{https://github.com/chroma-core/chroma}}.
The embeddings are generated from a smaller, pre-trained model\footnote{all-MiniLM-L6-v2 model with SentenceTransformaters \url{https://sbert.net/}} for better speed at a lower cost. When a query is issued, the system
retrieves the top $k$ most relevant embeddings from the vector database. 

Once the query results are rendered, they can optionally be processed further by the Context Graph Engine. This component can rehydrate the retrieved nodes into a larger subgraph, analyze the relationships between them,
and apply importance scoring to highlight key elements within the relevant context.

\section{Technical Innovations}
\vspace{0.3em} 

HCGS incorporates several key technical innovations that advance the state of the art in code understanding and retrieval. These innovations span three primary areas: representation enhancements, processing improvements, and retrieval advancements.

\begin{itemize}
  \item \textbf{Language-Agnostic Code Analysis:} The use of the Language Server Protocol (LSP) enables robust, multi-language code analysis without requiring language-specific parsers. This makes HCGS applicable to a wide range of programming languages and environments.

  \item \textbf{Hierarchical Context Propagation:} The bottom-up traversal strategy in Section \ref{sec:summary_generator} ensures that higher-level summaries incorporate context from all their dependencies, creating a rich, multi-layered understanding of code relationships that captures both low-level implementation details and high-level architectural patterns.

  \item \textbf{Context-Aware Embeddings:} Unlike traditional approaches that generate embeddings for functions in isolation, HCGS generates embeddings that incorporate contextual information from called functions. This results in significantly higher information density and semantic accuracy, addressing a fundamental limitation in code representation by capturing a function's role within the broader system architecture.

  \item \textbf{Structured Summary Schema:} The schema shown in Figure~\ref{fig:schema} organizes code information into well-defined categories, including implementation details and dependencies. This structured approach facilitates consistent documentation and enables sophisticated, targeted queries.

  \item \textbf{Context Graph Engine:} The innovative approach to result composition through the Context Graph Engine creates coherent subgraphs that represent relevant portions of the codebase. These subgraphs provide comprehensive context, including both directly relevant code elements and their important relationships.

\end{itemize}

Together, these innovations create a system that advances code understanding and retrieval capabilities.

\input{methodology_section}

\section{Limitations}
\label{sec:limitations}
\vspace{0.3em} 

While HCGS advances code understanding and retrieval, several inherent limitations warrant consideration. The system's reliance on static analysis techniques constrains its ability to capture dynamic behaviors resolved only during execution. Reflection-based invocations, dynamic dispatch patterns, and runtime-determined control flows remain challenging to incorporate, particularly in languages with extensive metaprogramming capabilities where execution paths may diverge significantly from static analysis predictions.

The hierarchical summaries' quality depends on the underlying Large Language Model's capabilities. Despite using state-of-the-art models, the summarization process remains susceptible to misinterpretations of complex code patterns, inconsistent abstraction levels, and occasional hallucinations with unfamiliar programming paradigms. These limitations can propagate through the hierarchy, affecting higher-level summaries that incorporate information from lower levels.

Although the Language Server Protocol enhances language coverage, integration challenges persist for new programming languages. Each language requires a robust LSP server supporting the necessary static analysis operations. Languages with less mature tooling ecosystems or specialized semantics may present difficulties, potentially limiting applicability in certain domains. Variations in LSP implementations can also lead to inconsistencies in code graph quality across languages.

A significant operational limitation is HCGS's approach to codebase updates, requiring complete re-parsing when changes occur. This process becomes computationally intensive for large projects, particularly problematic in dynamic development environments with frequent changes. Developing efficient incremental update mechanisms that selectively reprocess only affected portions of the code graph represents an important area for future research to improve responsiveness in real-world scenarios.

\section{Conclusion}
\vspace{0.3em} 

We presented Hierarchical Code Graph Summarization (HCGS), a novel system that enhances code retrieval through bottom-up analysis of code relationships. By leveraging the Language Server Protocol and structured summarization, HCGS achieves significant improvements in retrieval accuracy, demonstrating up to 82\% relative improvement in top-1 precision for large codebases and perfect Pass@3 scores for smaller repositories. The system's language-agnostic approach and context-aware embeddings address fundamental limitations in existing code retrieval systems, while future work will focus on incremental graph updates and enhanced runtime analysis. These advancements position HCGS as a significant step toward more effective code comprehension tools.

\bibliographystyle{plain} 
\bibliography{references.bib} 

\end{document}

%% file: methodology_section.tex
\section{Methodology}
\label{sec:methodology}

In this section, we present our methodology for evaluating the effectiveness of Hierarchical Code Graph Summarization (HCGS) in code retrieval tasks, addressing fundamental limitations in existing benchmarks while providing a rigorous framework for assessing the impact of hierarchical context on retrieval performance.

\subsection{Limitations of Existing Benchmarks}

Contemporary code retrieval benchmarks such as COIR \cite{coir} and CodeSearchNet \cite{husain2020codesearchnetchallengeevaluatingstate} exhibit several critical limitations that constrain their real-world applicability. As noted by Voyage AI \cite{voyage}, an estimated 51\% of labels in widely-used benchmarks like CoSQA are incorrect, with queries and code snippets being mismatched. Furthermore, existing benchmarks evaluate retrieval performance on isolated functions, failing to capture the intricate relationships between functions in real-world codebases. Current approaches typically comprise individual function definitions void of project structure, constrained only by programming language. Such search spaces omit essential contextual information present in most software engineering scenarios, neglecting the crucial relationships between code elements that are essential for comprehensive code comprehension.

\subsection{Evaluation Approach and Query Generation}

Our evaluation methodology compares two fundamental retrieval approaches: code-only retrieval using embeddings from raw code without hierarchical context, and summary-enhanced retrieval utilizing embeddings from hierarchical summaries that incorporate function dependency context. Unlike existing benchmarks, our approach includes complete codebases in the search space with multiple programming languages, preserving structural and contextual information typically lost in traditional benchmarks.

We developed a synthetic query generation pipeline leveraging Large Language Models to create realistic search queries based on function summaries. The process begins with generating structured summaries for each function using our hierarchical approach described in Section \ref{sec:summary_generator}, followed by condensing these summaries while preserving critical technical details. Using these condensed summaries, we generate multiple search queries (limited to 10 words) that focus on the function's purpose and behavior, producing queries that mirror how developers might search for code. 

\subsection{Benchmark Dataset Construction}

To ensure comprehensive evaluation across diverse scenarios, we selected five codebases of varying sizes and domains from trending GitHub repositories, collectively including functions written in multiple programming languages. Table \ref{tab:dataset_characteristics} presents the characteristics of these repositories, ranging from large codebases like libsignal (5,128 functions) to smaller ones like DeepSeek-V3 (44 functions). For each codebase, we generated a code graph capturing function dependencies, created hierarchical summaries, generated synthetic queries, and created relevance annotations. In total, our benchmark includes 6,594 queries across all five codebases with 7,531 functions, providing statistically significant evaluation across different scales and complexities.

\begin{table}[t]
\centering
\caption{Benchmark Dataset Characteristics: Five diverse codebases spanning multiple programming languages, domains, and sizes, providing a comprehensive test bed for evaluating code retrieval approaches.}
\label{tab:dataset_characteristics}
\vspace{5pt}
\begin{tabular}{lllrr}
\hline
\textbf{Repository} & \textbf{Domain} & \textbf{Primary Languages} & \textbf{Functions} & \textbf{Queries} \\
\hline
libsignal & Cryptography & C++, Java, Rust & 5,128 & 3,916 \\
ingress-nginx & Infrastructure & Go, Python & 1,773 & 1,965 \\
awesome-llm-apps & AI Applications & Python, JavaScript & 330 & 373 \\
newsnow & News Aggregation & TypeScript, JavaScript & 256 & 306 \\
DeepSeek-V3 & Deep Learning & Python & 44 & 34 \\
\hline
\textbf{Total} & & & \textbf{7,531} & \textbf{6,594} \\
\hline
\end{tabular}
\end{table}

\subsection{Evaluation Process and Metrics}

For each codebase, we created two vector database collections: one containing embeddings of raw function code and another with embeddings of hierarchical function summaries. For each query, we generated vector embeddings and performed retrieval against both databases, recording the top-k results (k = 1, 3, 5, 10). We calculated Pass@k (percentage of queries with a relevant function in top-k results), Coverage (percentage of annotated functions appearing in results), and NDCG (Normalized Discounted Cumulative Gain for ranking quality). The Pass@k metric reflects practical utility from a developer's perspective, with higher Pass@1 values indicating the system is more likely to return the most relevant function first. To ensure fair comparison, we used identical query processing and embedding generation techniques for both approaches, isolating the specific impact of hierarchical summarization.

\subsection{Results}

Tables \ref{tab:retrieval_results} and \ref{tab:absolute_results} present the comparative performance of summary-based and code-only retrieval across all codebases, demonstrating significant advantages of hierarchical code summarization for code retrieval.

\begin{table}[t]
    \centering
    \caption{Comparison of Hierarchical Code Summary vs Code Only Metrics}
    \label{tab:retrieval_results}
    \vspace{5pt}
    \begin{tabular}{l|rr|c|c|c|c|c|c|c}
    \hline
    & & & \multicolumn{2}{c|}{Pass@1} & \multicolumn{2}{c|}{Pass@3} & \multicolumn{2}{c}{Pass@10} \\
    \cline{4-9}
    Codebase & Queries & Functions & Abs.Imp & Rel.Imp & Abs.Imp & Rel.Imp & Abs.Imp & Rel.Imp \\
    \hline
    libsignal & 3916 & 5128 & +27.15 & +82\% & +29.16 & +56\% & +22.63 & +33\% \\
    ingress-nginx & 1965 & 1773 & +25.55 & +76\% & +29.38 & +57\% & +25.09 & +37\% \\
    awesome-llm-apps & 373 & 330 & +8.99 & +13\% & +5.45 & +6\% & +0.82 & +1\% \\
    newsnow & 306 & 256 & +11.44 & +19\% & +7.19 & +8\% & +2.61 & +3\% \\
    DeepSeek-V3 & 34 & 44 & +8.82 & +12\% & +5.88 & +6\% & +2.94 & +3\% \\
    \hline
    \end{tabular}
    \end{table}


\begin{table}[!htbp]
\centering
\caption{Absolute Values for Hierarchical Code Summary vs Code Only Metrics}
\label{tab:absolute_results}
\vspace{5pt}
\begin{tabular}{l|c|c|c|c|c|c}
\hline
& \multicolumn{2}{c|}{Pass@1} & \multicolumn{2}{c|}{Pass@3} & \multicolumn{2}{c}{Pass@10} \\
\cline{2-7}
Codebase & Code & Summary & Code & Summary & Code & Summary \\
\hline
libsignal & 33.07 & 60.21 & 51.63 & 80.80 & 69.41 & 92.03 \\
ingress-nginx & 33.73 & 59.27 & 51.66 & 81.04 & 67.50 & 92.59 \\
awesome-llm-apps & 68.12 & 77.11 & 89.10 & 94.55 & 98.64 & 99.46 \\
newsnow & 60.78 & 72.22 & 85.62 & 92.81 & 95.42 & 98.04 \\
DeepSeek-V3 & 73.53 & 82.35 & 94.12 & 100.00 & 97.06 & 100.00 \\
\hline
\end{tabular}
\end{table}

Our evaluation reveals that hierarchical summarization consistently outperforms code-only retrieval across all codebases and metrics, with both substantial absolute and relative improvements.

The improvement is most pronounced for larger, more complex codebases such as libsignal and ingress-nginx. For libsignal, the Pass@1 metric improved by 27.15 percentage points, representing an 82\% relative improvement over the baseline (from 33.07\% to 60.21\%). Similarly, ingress-nginx showed a 25.55 percentage point improvement in Pass@1, corresponding to a 76\% relative gain (from 33.73\% to 59.27\%). These substantial improvements in Pass@1 are particularly valuable in practical scenarios where developers typically focus on the top result.

The gains extend to Pass@3 metrics, where both large codebases showed improvements exceeding 29 percentage points (56-57\% relative improvement), suggesting that hierarchical context significantly enhances the quality of top results. Even at Pass@10, the improvements remain substantial, with absolute gains of 22-25 percentage points (33-37\% relative improvement) for these larger codebases.

For smaller codebases, while the absolute improvements are more modest, they still show consistent gains. The awesome-llm-apps repository demonstrated improvements of 8.99 percentage points in Pass@1 (13\% relative improvement), while newsnow showed an 11.44 percentage point gain (19\% relative improvement). The smallest codebase, DeepSeek-V3, achieved perfect Pass@3 and Pass@10 scores with the summary-based approach, representing improvements of 5.88 and 2.94 percentage points respectively (6\% and 3\% relative gains).

These results align with our hypothesis that hierarchical context provides greater benefits in larger, more interconnected codebases where understanding function relationships is crucial for accurate retrieval. The consistent pattern of improvements across all metrics and codebases, particularly the substantial relative gains in Pass@1 performance for larger codebases, demonstrates the effectiveness of our hierarchical summarization approach in enhancing code retrieval accuracy.